\documentstyle[aps,preprint]{revtex}

\begin{document}
\author{Remo Garattini}
\address{M\'{e}canique et Gravitation, Universit\'{e} de Mons-Hainaut,\\
Facult\'e des Sciences, 15 Avenue Maistriau, \\
B-7000 Mons, Belgium \\
and\\
Facolt\`a di Ingegneria, Universit\`a degli Studi di Bergamo,\\
Viale Marconi, 5, 24044 Dalmine (Bergamo) Italy\\
e-mail: Garattini@mi.infn.it}
\title{Entropy from the foam}
\date{\today}
\maketitle

\begin{abstract}
A simple model of spacetime foam, made by $N$ wormholes in a semiclassical
approximation, is taken under examination. We show that the
Bekenstein-Hawking entropy is here {\it ``quantized''} in agreement with the
heuristic calculation of Bekenstein.
\end{abstract}

\section{Introduction}

In the early seventies J. Bekenstein\cite{J.Bekenstein} and S. Hawking\cite
{S.Hawking}, basing themselves upon considerations originated from quantum
field theory, changed drastically the concept of a black hole. Beginning
with the simple observation that the area of the horizon of the
Schwarzschild black hole is a quadratic function of the mass $M$, they
considered an infinitesimal increment (in natural units $G=\hbar =c=1$) 
\[
dM=\kappa dA_{hor}\text{\qquad }\kappa =\frac{1}{4M}.
\]
This formally resembles the First Law of thermodynamics $dU=TdS$\footnote{%
The best way to see this similarity is with a Kerr-Newman black hole, where
the incremental formula is 
\[
dM\equiv \kappa dA_{hor}+\Phi dQ+\vec{\Omega}\cdot d\vec{L}
\]
and the analogy with the First Law 
\[
dU=TdS+PdV
\]
is complete.}. This analogy of the black hole mechanics with thermodynamics
appears reinforced by Hawking's theorem on black hole mechanics\cite
{S.Hawking1} asserting that the horizon area of an isolated black hole never
decreases in any transformation. Thus, if two black holes of area $A_{1}$
and $A_{2}$ fuse to form a black hole of area $A_{1+2}$, then the Theorem
asserts that $A_{1+2}\geq A_{1}+A_{2}.$ On the basis of these observation
and results, Bekenstein made the proposal that a black hole {\it does} have
an entropy proportional to the area of its horizon 
\begin{equation}
S_{bh}=const\times A_{hor}.
\end{equation}
In particular, in natural units one finds that the proportionality constant
is set to $1/4G=1/4l_{p}^{2}$, so that the entropy becomes 
\begin{equation}
S=\frac{A}{4G}=\frac{A}{4l_{p}^{2}}.
\end{equation}
For a Schwarzschild black hole, for example, one finds the value 
\begin{equation}
S=\frac{4\pi \left( 2MG\right) ^{2}}{4G}=4\pi M^{2}G=4\pi M^{2}l_{p}^{2}.
\end{equation}
In conventional units instead, we find for a generic horizon area that 
\begin{equation}
S_{bh}=\frac{1}{8\pi G\hbar }\ln 2k_{B}c^{3}A_{hor}.
\end{equation}
This formula is identical (except for the factor of $\ln 2$ which one may
think of as a choice of units of entropy) to the one proposed by Hawking\cite
{S.Hawking} based on consistency with the rate of black hole radiation. The
appearance of $\hbar $ is ``... a reflection of the fact that the entropy is
... a count of states of the system''\cite{J.Bekenstein}. Following
Bekenstein's proposal on the quantization of the area for nonextremal black
holes we have 
\[
a_{n}=\alpha l_{p}^{2}\left( n+\eta \right) \text{\qquad }\eta >-1\text{%
\qquad }n=1,2,\ldots 
\]
Many attempts to recover the area spectrum have been done, see Refs.\cite
{Bekenstein,Mukohyama} for a review. Note that the appearance of a discrete
spectrum is not so trivial. Indeed there are other theories, based on
spherically symmetric metrics in a mini-superspace approach, whose mass
spectrum is continuous. Recently a model made by $N$ {\it coherent}
wormholes, based on Wheeler's ideas of a foamy spacetime\cite{Wheeler}, has
been considered\cite{Remo}. In that paper, hereafter referred as I, we have
computed the energy density of gravitational fluctuations reproducing the
same behavior conjectured by Wheeler during the sixties on dimensional
grounds. In this paper we wish to apply the ideas of \ I to a generic area.
The result is a {\it quantization} process whose quanta can be identified
with wormholes of Planckian size. Implications on the black hole entropy are
taken under consideration. The rest of the paper is structured as follows,
in section \ref{p2} we briefly recall the results reported in Ref.\cite{Remo}%
, in section \ref{p3} we compute the area spectrum. We summarize and
conclude in section \ref{p4}.

\section{Spacetime foam: the model}

\label{p2}In the one-wormhole approximation we have used an eternal black
hole, to describe a complete manifold ${\cal M}$, composed of two wedges $%
{\cal M}_{+}$ and ${\cal M}_{-}$ located in the right and left sectors of a
Kruskal diagram. The spatial slices $\Sigma $ represent Einstein-Rosen
bridges with wormhole topology $S^{2}\times R^{1}$. Also the hypersurface $%
\Sigma $ is divided in two parts $\Sigma _{+}$ and $\Sigma _{-}$ by a
bifurcation two-surface $S_{0}$. We begin with the line element 
\begin{equation}
ds^{2}=-N^{2}\left( r\right) dt^{2}+\frac{dr^{2}}{1-\frac{2m}{r}}%
+r^{2}\left( d\theta ^{2}+\sin ^{2}\theta d\phi ^{2}\right)   \label{a1}
\end{equation}
and we consider the physical Hamiltonian defined on $\Sigma $%
\[
H_{P}=H-H_{0}=\frac{1}{16\pi l_{p}^{2}}\int_{\Sigma }d^{3}x\left( N{\cal H}%
+N_{i}{\cal H}^{i}\right) 
\]
\begin{equation}
+\frac{2}{l_{p}^{2}}\int_{S_{+}}^{{}}d^{2}x\sqrt{\sigma }\left(
k-k^{0}\right) -\frac{2}{l_{p}^{2}}\int_{S_{-}}d^{2}x\sqrt{\sigma }\left(
k-k^{0}\right) ,
\end{equation}
where $l_{p}^{2}=G$. The volume term contains two constraints 
\begin{equation}
\left\{ 
\begin{array}{l}
{\cal H}=G_{ijkl}\pi ^{ij}\pi ^{kl}\left( \frac{l_{p}^{2}}{\sqrt{g}}\right)
-\left( \frac{\sqrt{g}}{l_{p}^{2}}\right) R^{\left( 3\right) }=0 \\ 
{\cal H}^{i}=-2\pi _{|j}^{ij}=0
\end{array}
\right. ,  \label{a1a}
\end{equation}
both satisfied by the Schwarzschild and Flat metric respectively. The
supermetric is $G_{ijkl}=\frac{1}{2}\left(
g_{ik}g_{jl}+g_{il}g_{jk}-g_{ij}g_{kl}\right) $ and $R^{\left( 3\right) }$
denotes the scalar curvature of the surface $\Sigma $. By using the
expression of the trace 
\begin{equation}
k=-\frac{1}{\sqrt{h}}\left( \sqrt{h}n^{\mu }\right) _{,\mu },
\end{equation}
with the normal to the boundaries defined continuously along $\Sigma $ as $%
n^{\mu }=\left( h^{yy}\right) ^{\frac{1}{2}}\delta _{y}^{\mu }$. The value
of $k$ depends on the function $r,_{y}$, where we have assumed that the
function $r,_{y}$ is positive for $S_{+}$ and negative for $S_{-}$. We
obtain at either boundary that 
\begin{equation}
k=\frac{-2r,_{y}}{r}.
\end{equation}
The trace associated with the subtraction term is taken to be $k^{0}=-2/r$
for $B_{+}$ and $k^{0}=2/r$ for $B_{-}$. Then the quasilocal energy with
subtraction terms included is 
\begin{equation}
E_{{\rm quasilocal}}=l_{p}^{2}\left( E_{+}-E_{-}\right) =l_{p}^{2}\left[
\left( r\left[ 1-\left| r,_{y}\right| \right] \right) _{y=y_{+}}-\left( r%
\left[ 1-\left| r,_{y}\right| \right] \right) _{y=y_{-}}\right] .
\end{equation}
Note that the total quasilocal energy is zero for boundary conditions
symmetric with respect to the bifurcation surface $S_{0}$ and this is the
necessary condition to obtain instability with respect to the flat space. In
this sector satisfy the constraint equations (\ref{a1a}). Here we consider
perturbations at $\Sigma $ of the type 
\begin{equation}
g_{ij}=\bar{g}_{ij}+h_{ij},
\end{equation}
where $\bar{g}_{ij}$ is the spatial part of the Schwarzschild and Flat
background in a WKB approximation. In this framework we have computed the
quantity 
\begin{equation}
\Delta E\left( M\right) =\frac{\left\langle \Psi \left|
H^{Schw.}-H^{Flat}\right| \Psi \right\rangle }{\left\langle \Psi |\Psi
\right\rangle }+\frac{\left\langle \Psi \left| H_{quasilocal}\right| \Psi
\right\rangle }{\left\langle \Psi |\Psi \right\rangle },  \label{a2}
\end{equation}
by means of a variational approach, where the WKB functionals are
substituted with trial wave functionals. This quantity is the natural
extension to the volume term of the subtraction procedure for boundary terms
and it is interpreted as the Casimir energy related to vacuum fluctuations.
By restricting our attention to the graviton sector of the Hamiltonian
approximated to second order, hereafter referred as $H_{|2}$, we define 
\[
E_{|2}=\frac{\left\langle \Psi ^{\perp }\left| H_{|2}\right| \Psi ^{\perp
}\right\rangle }{\left\langle \Psi ^{\perp }|\Psi ^{\perp }\right\rangle },
\]
where 
\[
\Psi ^{\perp }=\Psi \left[ h_{ij}^{\perp }\right] ={\cal N}\exp \left\{ -%
\frac{1}{4l_{p}^{2}}\left[ \left\langle \left( g-\bar{g}\right) K^{-1}\left(
g-\bar{g}\right) \right\rangle _{x,y}^{\perp }\right] \right\} .
\]
After having functionally integrated $H_{|2}$, we get 
\begin{equation}
H_{|2}=\frac{1}{4l_{p}^{2}}\int_{\Sigma }d^{3}x\sqrt{g}G^{ijkl}\left[
K^{-1\bot }\left( x,x\right) _{ijkl}+\left( \triangle _{2}\right)
_{j}^{a}K^{\bot }\left( x,x\right) _{iakl}\right] 
\end{equation}
The propagator $K^{\bot }\left( x,x\right) _{iakl}$ comes from a functional
integration and it can be represented as 
\begin{equation}
K^{\bot }\left( \overrightarrow{x},\overrightarrow{y}\right)
_{iakl}:=\sum_{N}\frac{h_{ia}^{\bot }\left( \overrightarrow{x}\right)
h_{kl}^{\bot }\left( \overrightarrow{y}\right) }{2\lambda _{N}\left(
p\right) },
\end{equation}
where $h_{ia}^{\bot }\left( \overrightarrow{x}\right) $ are the
eigenfunctions of 
\begin{equation}
\left( \triangle _{2}\right) _{j}^{a}:=-\triangle \delta
_{j}^{a_{{}}^{{}}}+2R_{j}^{a}.
\end{equation}
This is the Lichnerowicz operator projected on $\Sigma $ acting on traceless
transverse quantum fluctuations and $\lambda _{N}\left( p\right) $ are
infinite variational parameters. $\triangle $ is the curved Laplacian
(Laplace-Beltrami operator) on a Schwarzschild background and $R_{j\text{ }%
}^{a}$ is the mixed Ricci tensor whose components are:

\begin{equation}
R_{j}^{a}=diag\left\{ \frac{-2MG}{r_{{}}^{3}},\frac{MG}{r_{{}}^{3}},\frac{MG%
}{r_{{}}^{3}}\right\} .
\end{equation}
The minimization with respect to $\lambda $ and the introduction of a high
energy cutoff $\Lambda $ give to the Eq. (\ref{a2}) the following form 
\begin{equation}
\Delta E\left( M\right) \sim -\frac{V}{32\pi ^{2}}\left( \frac{3MG}{r_{0}^{3}%
}\right) ^{2}\ln \left( \frac{r_{0}^{3}\Lambda ^{2}}{3MG}\right) ,
\label{a3}
\end{equation}
where $V$ is the volume of the system and $r_{0}$ is related to the minimum
radius compatible with the wormhole throat. We know that the classical
minimum is achieved when $r_{0}=2MG$. However, it is likely that quantum
processes come into play at short distances, where the wormhole throat is
defined, introducing a {\it quantum} radius $r_{0}>2MG$. We now compute the
minimum of $\Delta E\left( M\right) $, after having rescaled the variable $M$
to a scale variable $x=3MG/\left( r_{0}^{3}\Lambda ^{2}\right) $. Thus 
\[
\Delta E\left( M\right) \rightarrow \Delta E\left( x,\Lambda \right) =\frac{V%
}{32\pi ^{2}}\Lambda ^{4}x^{2}\ln x 
\]
We obtain two values for $x$: $x_{1}=0$, i.e. flat space and $x_{2}=e^{-%
\frac{1}{2}}$. At the minimum 
\begin{equation}
\Delta E\left( x_{2}\right) =-\frac{V}{64\pi ^{2}}\frac{\Lambda ^{4}}{e}.
\end{equation}
Nevertheless, there exists another part of the spectrum which has to be
considered: the discrete spectrum containing one mode. This gives the energy
an imaginary contribution, namely we are discovered an unstable mode\cite
{GPY,Remo1}. Let us briefly recall, how this appears. The eigenvalue
equation 
\begin{equation}
\left( \triangle _{2}\right) _{i}^{a}h_{aj}=\alpha h_{ij}
\end{equation}
can be studied with the Regge-Wheeler method. The perturbations can be
divided in odd and even components. The appearance of the unstable mode is
governed by the gravitational field component $h_{11}^{even}$. Explicitly 
\[
-E^{2}H\left( r\right) 
\]
\begin{equation}
=-\left( 1-\frac{2MG}{r}\right) \frac{d^{2}H\left( r\right) }{dr^{2}}+\left( 
\frac{2r-3MG}{r^{2}}\right) \frac{dH\left( r\right) }{dr}-\frac{4MG}{r^{3}}%
H\left( r\right) ,  \label{a4}
\end{equation}
where 
\begin{equation}
h_{11}^{even}\left( r,\vartheta ,\phi \right) =\left[ H\left( r\right)
\left( 1-\frac{2m}{r}\right) ^{-1}\right] Y_{00}\left( \vartheta ,\phi
\right)
\end{equation}
and $E^{2}>0$. Eq. (\ref{a4}) can be transformed into 
\begin{equation}
\mu =\frac{\int\limits_{0}^{\bar{y}}dy\left[ \left( \frac{dh\left( y\right) 
}{dy}\right) ^{2}-\frac{3}{2\rho \left( y\right) ^{3}}h\left( y\right) %
\right] }{\int\limits_{0}^{\bar{y}}dyh^{2}\left( y\right) },
\end{equation}
where $\mu $ is the eigenvalue, $y$ is the proper distance from the throat
in dimensionless form. If we choose $h\left( \lambda ,y\right) =\exp \left(
-\lambda y\right) $ as a trial function we numerically obtain $\mu =-.701626$%
. In terms of the energy square we have 
\begin{equation}
E^{2}=-.\,\allowbreak 175\,41/\left( MG\right) ^{2}
\end{equation}
to be compared with the value $E^{2}=-.\,\allowbreak 19/\left( MG\right)
^{2} $ of Ref.\cite{GPY}. Nevertheless, when we compute the eigenvalue as a
function of the distance $y$, we discover that in the limit $\bar{y}%
\rightarrow 0$, 
\begin{equation}
\mu \equiv \ \mu \left( \lambda \right) =\lambda ^{2}-\frac{3}{2}+\frac{9}{8}%
\left[ \bar{y}^{2}+\frac{\bar{y}}{2\lambda }\right] .
\end{equation}
Its minimum is at $\tilde{\lambda}=\left( \frac{9}{32}\bar{y}\right) ^{\frac{%
1}{3}}$ and 
\begin{equation}
\mu \left( \tilde{\lambda}\right) =1.\,\allowbreak 287\,8\bar{y}^{\frac{2}{3}%
}+\frac{9}{8}\bar{y}^{2}-\frac{3}{2}.
\end{equation}
It is evident that there exists a critical radius where $\mu $ turns from
negative to positive. This critical value is located at $\rho
_{c}=1.\,\allowbreak 113\,4$ to be compared with the value $\rho _{c}=1.445$
obtained by B. Allen in \cite{B.Allen}. What is the relation with the large
number of wormholes? As mentioned in I, when the number of wormholes grows,
to keep the coherency assumption valid, the space available for every single
wormhole has to be reduced to avoid overlapping of the wave functions. If we
fix the initial boundary at $R_{\pm }$, then in presence of $N_{w}$
wormholes, it will be reduced to $R_{\pm }/N_{w}$. This means that boundary
conditions are not fixed at infinity, but at a certain finite radius and the 
$ADM$ mass term is substituted by the quasilocal energy expression under the
condition of having symmetry with respect to each bifurcation surface. The
effect on the unstable mode is clear: as $N_{w}$ grows, the boundary radius
reduces more and more until it will reach the critical value $\rho _{c}$
below which no negative mode will appear corresponding to a critical
wormholes number $N_{w_{c}}$. To this purpose, suppose to consider $N_{w}$
wormholes and assume that there exists a covering of $\Sigma $ such that $%
\Sigma =\bigcup\limits_{i=1}^{N_{w}}\Sigma _{i}$, with $\Sigma _{i}\cap
\Sigma _{j}=\emptyset $ when $i\neq j$. Each $\Sigma _{i}$ has the topology $%
S^{2}\times R^{1}$ with boundaries $\partial \Sigma _{i}^{\pm }$ with
respect to each bifurcation surface. On each surface $\Sigma _{i}$,
quasilocal energy gives 
\begin{equation}
E_{i\text{ }{\rm quasilocal}}=\frac{2}{l_{p}^{2}}\int_{S_{i+}}d^{2}x\sqrt{%
\sigma }\left( k-k^{0}\right) -\frac{2}{l_{p}^{2}}\int_{S_{i-}}d^{2}x\sqrt{%
\sigma }\left( k-k^{0}\right) .
\end{equation}
Thus if we apply the same procedure of the single case on each wormhole, we
obtain 
\begin{equation}
E_{i\text{ }{\rm quasilocal}}=l_{p}^{2}\left( E_{i+}-E_{i-}\right)
=l_{p}^{2}\left( r\left[ 1-\left| r,_{y}\right| \right] \right)
_{y=y_{i+}}-l_{p}^{2}\left( r\left[ 1-\left| r,_{y}\right| \right] \right)
_{y=y_{i-}}.
\end{equation}
Note that the total quasilocal energy is zero for boundary conditions
symmetric with respect to {\it each} bifurcation surface $S_{0,i}$. We are
interested in a large number of wormholes, each of them contributing with a
term of the type (\ref{a2}). If the wormholes number is $N_{w}$, we obtain
(semiclassically, i.e., without self-interactions)\footnote{%
Note that at this approximation level, we are in the same situation of a
large collection of N harmonic oscillators whose hamiltonian is 
\[
H=\frac{1}{2}\sum_{n\neq 0}^{\infty }\left[ \pi _{n}^{2}+n^{2}\omega
^{2}\phi _{n}^{2}\right] . 
\]
} 
\begin{equation}
H_{tot}^{N_{w}}=\underbrace{H^{1}+H^{2}+\ldots +H^{N_{w}}}.
\end{equation}
Thus the total energy for the collection is 
\[
E_{|2}^{tot}=N_{w}H_{|2}. 
\]
The same happens for the trial wave functional which is the product of $%
N_{w} $ t.w.f.. Thus 
\[
\Psi _{tot}^{\perp }=\Psi _{1}^{\perp }\otimes \Psi _{2}^{\perp }\otimes
\ldots \ldots \Psi _{N_{w}}^{\perp }={\cal N}\exp N_{w}\left\{ -\frac{1}{%
4l_{p}^{2}}\left[ \left\langle \left( g-\bar{g}\right) K^{-1}\left( g-\bar{g}%
\right) \right\rangle _{x,y}^{\perp }\right] \right\} 
\]
\[
={\cal N}\exp \left\{ -\frac{1}{4}\left[ \left\langle \left( g-\bar{g}%
\right) K^{-1}\left( g-\bar{g}\right) \right\rangle _{x,y}^{\perp }\right]
\right\} , 
\]
where we have rescaled the fluctuations $h=g-\bar{g}$ in such a way to
absorb $N_{w}/l_{p}^{2}$. The propagator $K^{\bot }\left( x,x\right) _{iakl}$
is the same one \ for the one wormhole case. Thus, repeating the same steps
of the single wormhole, but in the case of $N_{w}$ wormholes, one gets 
\begin{equation}
\Delta E_{N_{w}}\left( x,\Lambda \right) \sim N_{w}^{2}\frac{V}{32\pi ^{2}}%
\Lambda ^{4}x^{2}\ln x,
\end{equation}
where we have defined the usual scale variable $x=3MG/\left(
r_{0}^{3}\Lambda ^{2}\right) $. Then at one loop the cooperative effects of
wormholes behave as one {\it macroscopic single }field multiplied by $%
N_{w}^{2}$, but without the unstable mode. At the minimum, $\bar{x}=e^{-%
\frac{1}{2}}$ 
\begin{equation}
\Delta E\left( \bar{x}\right) =-N_{w}^{2}\frac{V}{64\pi ^{2}}\frac{\Lambda
^{4}}{e}.
\end{equation}

\section{Area spectrum and Entropy}

\label{p3}A very important application of the model presented in the
previous section is the area quantization. The area is measured by the
quantity 
\begin{equation}
A\left( S_{0}\right) =\int_{S_{0}}d^{2}x\sqrt{\sigma }.
\end{equation}
$\sigma $ is the two-dimensional determinant coming from the induced metric $%
\sigma _{ab}$ on the boundary $S_{0}$. We would like to evaluate the mean
value of the area 
\begin{equation}
A\left( S_{0}\right) =\frac{\left\langle \Psi _{F}\left| \hat{A}\right| \Psi
_{F}\right\rangle }{\left\langle \Psi _{F}|\Psi _{F}\right\rangle }=\frac{%
\left\langle \Psi _{F}\left| \widehat{\int_{S_{0}}d^{2}x\sqrt{\sigma }}%
\right| \Psi _{F}\right\rangle }{\left\langle \Psi _{F}|\Psi
_{F}\right\rangle },
\end{equation}
computed on 
\begin{equation}
\left| \Psi _{F}\right\rangle =\Psi _{1}^{\perp }\otimes \Psi _{2}^{\perp
}\otimes \ldots \ldots \Psi _{N_{w}}^{\perp }.
\end{equation}
If we use the fact that $\sigma _{ab}=\bar{\sigma}_{ab}+\delta \sigma _{ab}$%
, where $\bar{\sigma}_{ab}$ is such that $\int_{S_{0}}d^{2}x\sqrt{\bar{\sigma%
}}=4\pi \bar{r}^{2}$ and $\bar{r}$ is the radius of $S_{0}$, we obtain to
the lowest level\footnote{%
For lowest level in the expansion of $\sigma _{ab}$, we mean 
\[
\sqrt{\sigma _{ab}}=\exp Tr\ln \sqrt{\sigma _{ab}}=\exp Tr\frac{1}{2}\ln
\left( \bar{\sigma}_{ab}+\delta \sigma _{ab}\right) 
\]
\[
=\exp Tr\frac{1}{2}\left[ \ln \bar{\sigma}_{ab}\left( 1+\frac{\delta \sigma
_{ab}}{\bar{\sigma}_{ab}}\right) \right] =\exp Tr\frac{1}{2}\left[ \ln \bar{%
\sigma}_{ab}+\ln \left( 1+\frac{\delta \sigma _{ab}}{\bar{\sigma}_{ab}}%
\right) \right] \simeq 
\]
\begin{equation}
\sqrt{\bar{\sigma}_{ab}}+o\left( \delta \sigma _{ab}\right) 
\end{equation}
} in the expansion of $\sigma _{ab}$ that 
\begin{equation}
A\left( S_{0}\right) =\frac{\left\langle \Psi _{F}\left| \hat{A}\right| \Psi
_{F}\right\rangle }{\left\langle \Psi _{F}|\Psi _{F}\right\rangle }=4\pi 
\bar{r}^{2}.  \label{p3a}
\end{equation}
Suppose to consider the mean value of the area $A$ computed on a given {\it %
macroscopic} fixed radius $R$. On the basis of our foam model, we obtain $%
A=\bigcup\limits_{i=1}^{N}A_{i}$, with $A_{i}\cap A_{j}=\emptyset $ when $%
i\neq j$. Thus 
\begin{equation}
A=4\pi R^{2}=\sum\limits_{i=1}^{N}A_{i}=\sum\limits_{i=1}^{N}4\pi \bar{r}%
_{i}^{2}.
\end{equation}
When $\bar{r}_{i}\rightarrow l_{p}$, $A_{i}\rightarrow A_{l_{P}}$ and 
\begin{equation}
A=NA_{l_{P}}=N4\pi l_{p}^{2}.  \label{p3b}
\end{equation}
Thus the {\it macroscopic} area is represented by $N$ {\it microscopic}
areas of the Planckian size: in this sense we will claim that the area is
quantized. One immediately observes that $N$ is such that $N\geq N_{w_{c}}$,
where $N_{w_{c}}$ is the critical wormholes number above which we have the
stability of our foam model. At this point we can apply the same reasoning
of Refs.\cite{Rovelli,Rovelli1,Fabio} to arrive at the Bekenstein-Hawking
relation between entropy and area 
\begin{equation}
S=\frac{A}{4l_{p}^{2}}.  \label{p3c}
\end{equation}
Note that there is a $\ln 2$ numerical factor missing to complete the
equality, between this model and the models described in Refs.\cite
{Rovelli,Rovelli1,Fabio}. This is principally due to the fact that there is
no a degeneracy factor related to the statistics. Here the wormholes are the
same. Of course the introduction of a degeneracy factor does not alter the
main result and the only effect will be that of dividing the spacetime
covering into equivalence classes with a representative for each class. In
our case, we deduce that the entropy is 
\begin{equation}
S=N\pi .
\end{equation}
Moreover this seems also to agree with the conclusions of Bekenstein (\cite
{Bekenstein} and Refs. therein), apart the degeneracy factor that in our
model seems to be related to the odd or even permutation of any single
wormhole wavefunction\footnote{{\bf Remark. }Entropy is {\it ``quantized''}
as a{\bf \ }consequence of $\left( \ref{p3b}\right) $ and not as a direct
application of the definition $S=-\sum p_{n}\ln p_{n}.$}. We can use Eq. (%
\ref{p3b}) to compute the entropy also for other geometries, for example,
the de Sitter geometry. Since we know that for this metric the
Bekenstein-Hawking relation (\ref{p3c}) still holds, we write 
\begin{equation}
S=\frac{3\pi }{l_{p}^{2}\Lambda }=\frac{A}{4l_{p}^{2}}=\frac{N4\pi l_{p}^{2}%
}{4l_{p}^{2}}=N\pi ,
\end{equation}
that is\footnote{%
A relation relating $\Lambda $ and $G$, via an integer $N$ appeared also in
Ref.\cite{Nojiri}. Nevertheless in Ref.\cite{Nojiri}, $N$ represents the
number of scalar fields and the bound from above and below 
\[
\left| 2GN\Lambda /3-2\right| \geq \sqrt{3}
\]
comes into play, instead of the equality $\left( \ref{p3d}\right) .$} 
\begin{equation}
\frac{3}{l_{p}^{2}N}=\Lambda .  \label{p3d}
\end{equation}
Thus the cosmological constant $\Lambda $ is ``{\it quantized''} in terms of 
$l_{p}$. Note that when the wormholes number $N$ is quite ``{\it large}'', $%
\Lambda \rightarrow 0.$ We could try to see what is the rate of change
between an early universe value of the cosmological constant and the value
that we observe. In inflationary models of the early universe is assumed to
have undergone an early phase with a large effective $\Lambda \sim \left(
10^{10}-10^{11}GeV\right) ^{2}$ for GUT era inflation, or $\Lambda \sim
\left( 10^{16}-10^{18}GeV\right) ^{2}$ for Planck era inflation. A
subsequent phase transition would then produce a region of space-time with $%
\Lambda \leq \left( 10^{-42}GeV\right) ^{2}$, i.e. the space in which we now
live. For GUT era inflation, we have (we are looking only at the order of
magnitude) 
\begin{equation}
10^{20}-10^{22}GeV^{2}=\frac{1}{N}10^{38}GeV^{2}\rightarrow
N=10^{16}-10^{18},
\end{equation}
while for Planck era inflation we have 
\begin{equation}
10^{32}-10^{36}GeV^{2}=\frac{1}{N}10^{38}GeV^{2}\rightarrow N=10^{6}-10^{2},
\end{equation}
to be compared with the value of $\left( 10^{-42}GeV\right) ^{2}$ which
gives a wormholes number of the order of 
\begin{equation}
10^{-84}GeV^{2}=\frac{1}{N}10^{38}GeV^{2}\rightarrow N=10^{122}.
\end{equation}
This very huge number is obtained by averaging the area on a cosmological
scale by means of Planck scale wormholes. Thus it seems quite reasonable
that with the growing of the cosmological radius, we obtain a growing \
wormholes number covering the horizon area.

\section{Conclusions}

\label{p4}

In this paper we have applied the model presented in I \ to the entropy
computation assuming the validity of the Bekenstein-Hawking relation. In
this picture the area is ``{\it quantized}'' in the sense that spacetime can
be filled by a given integer number of disjoint non-interacting wormholes.
This result is in agreement (apart a numerical factor) with the quantized
area proposed heuristically by Bekenstein and also with the loop quantum
gravity predictions of Refs.\cite{Rovelli,Rovelli1}, apart the degeneracy
factor missing, principally due to the fact that we have an ``{\it ideal
Boltzmann gas''} of wormholes. A similar result appeared in Ref. \cite{Fabio}%
. Nevertheless, in Ref.\cite{Fabio}, spacetime was assumed {\it ab initio}
built up of cells of Planckian size, while it seems that, in order to have
stability, spacetime needs to be covered by $N$ wormholes of the Planckian
size. Nevertheless, even if in this letter also the meaning of the
cosmological constant seems to be related to a pure gravitational effect
enforcing the idea that $\Lambda $ is not fundamental but is the effect of
quantum fluctuations of the pure gravitational field, giving therefore
strong indications of a foamy spacetime, we are not in presence of a model
which can cure the well known \ problems of the still absent theory of
quantum gravity. The main reason resides in a cutoff dependent model. Even
if one can argue that a Planck length cutoff is quite natural, it is not
clear how to compute such a value from first principles, even if it is
likely that a relation between this formulation of quantum gravity based on
a large number of coherent wormholes and the well accepted string theory
could exist.

\section{Acknowledgments}

I wish to thank R. Brout, M. Cavagli\`{a}, C. Kiefer, D. Hochberg, G.
Immirzi, S. Liberati, P. Spindel and M. Visser for useful comments and
discussions.

\end{document}